\documentclass{iau} 
\usepackage{graphicx}
\usepackage{color}

\title[Distances of Stars by mean of the Phase-lag Method] 
{Distances of Stars by mean of the Phase-lag Method}

\author[Sandra Etoka et al.]  
{Sandra Etoka$^{1,2}$,
 Dieter Engels$^2$,
 Eric {G\' erard}$^3$,
 Anita M.S. Richards$^1$}

\affiliation{$^1$Jodrell Bank Centre for Astrophysics, University of Manchester, UK
\\ email: {\tt Sandra.Etoka@googlemail.com} \\[\affilskip]
$^2$Hamburger Sternwarte, Universit\"{a}t Hamburg, Germany \\[\affilskip]
$^3$GEPI, Observatoire de Paris-Meudon, France}

\pubyear{2017}
\volume{336} 
\jname{Astrophysical Masers: Unlocking the Mysteries of the Universe}
\editors{A. Tarchi, M.J. Reid \& P. Castangia, eds.}
\begin{document}

\maketitle

\begin{abstract}
Variable OH/IR stars are Asymptotic Giant Branch (AGB) stars with an 
optically thick circumstellar envelope that emit strong OH 1612 MHz emission. 
They are commonly observed throughout the Galaxy but also in the LMC and
SMC. Hence, the precise inference of the distances of these stars will
ultimately result in better constraints on their mass range in
different metallicity environments. Through a multi-year long-term
monitoring program at the Nancay Radio telescope (NRT) and a
complementary high-sensitivity mapping campaign at the eMERLIN and
JVLA to measure precisely the angular diameter of the envelopes, we
have been re-exploring distance determination through the phase-lag
method for a sample of stars, in order to refine the poorly-constrained 
distances of some and infer the currently unknown distances of others. 
We present here an update of this project.

\keywords{masers, stars: late-type, stars: variables: OH/IR, stars: distances}
\end{abstract}

\firstsection 
\section{Introduction}
Evolved stars at the tip of the AGB for low- and in\-ter\-me\-diary-mass 
stars experience heavy mass loss surrounding the star with a circumstellar 
envelope (CSE), which ultimately become opaque to visible light. 
These enshrouded OH/IR stars 
commonly exhibit strong periodic (ranging typically from 1 to 6~yr) 
ground-state OH maser emission in the 1612-MHz transition.
Over 2000 OH masers of stellar origin are currently known in the Milky Way 
(\cite[Engels \& Bunzel, 2015]{Engels15b}) and it is anticipated that the SKA 
will detect thousands of OH maser sources of stellar origin in the anti-solar 
Galactic hemisphere and Local Group of galaxies 
(\cite[Etoka \etal\ 2015]{Etoka_etal15}).
This makes OH/IR stars potentially valuable objects for a wide range of studies 
in our Galaxy but also for stellar-evolution metallicity-related studies. \\

Because OH/IR stars are optically thick, their distances cannot be inferred 
using optical parallaxes. The period-luminosity relation found towards Miras 
(\cite[Whitelock, Feast \& Catchpole 1991]{Whitelock91}) breaks down for 
P$>$ 450~days. Kinematic distances can be very imprecise due 
to peculiar motions (\cite[Reid et al. 2009]{Reid09}). As it has been 
extensively demonstrated in this symposium, maser emission at high(er) 
frequency from e.g. water and methanol species is successfully used to 
infer distances via parallax measurements towards distant Galactic star 
forming regions. The use of ground-state OH maser to infer distances of 
AGB stars via parallax measurements has also been successfully done but 
only for objects relatively nearby (i.e., $\le$2~kpc; 
\cite[Vlemmings \& van Langevelde 2007]{Vlemmings07};
\cite[Orosz \etal\ 2017]{Orosz17}). 
Another alternative to distance determination for more distant evolved stars 
is the use of the ``phase-lag'' method.

\section{Method and Observations}\label{sec: Method}
The determination of the distance of an OH/IR star via the phase-lag method 
relies on the measurement of the linear and angular diameter of its 
OH-maser CSE which are both obtained independently.
OH/IR stars typically exhibit a double-peaked spectral profile where the 
blueshifted peak (``blue'' peak here after) emanates from the front cap of the 
CSE and the redshifted peak (``red'' peak here after) emanates from the back 
cap of the CSE while the faint interpeak emission emanates from the outer 
part of the CSE. 
We measure the phase lag ($\tau_{0}$) of a source with no external fitting 
function, using simply the shape of the light curve, by scaling and shifting 
the integrated-flux light curves of the blue peak $F_b$ with respect to the 
red one $F_r$, minimizing the function 
$\Delta F = F_{r}(t) - a \cdot F_{b}(t-\tau_0) + c$ 
(where $a$ and $c$ are constants for the amplitude and mean flux) leading to 
the measurement of the linear diameter of the OH shell of the star.
The angular diameter is obtained from interferometric mapping. \\

\cite[Schultz, Sherwood \& Winnberg (1978)]{Schultz78} performed the 
first phase-lag measurements, and in the 1980's, 
\cite[Herman \& Habing (1985)]{Herman85} and 
\cite[van Langevelde, van der Heiden \& van Schooneveld (1990)]{vanLangevelde_etal90}
explored this method to retrieve distances from OH/IR stars, but there are 
discrepancies in the phase-lag measurement of a good fraction of the sources
in common in these 2 works.
In an attempt to constrain the distance uncertainties achievable with this 
method we are re-exploring it. 
Our sample consists of 20 OH/IR stars that we have been monitoring 
with the {Nan\c cay} Radio Telescope (NRT) in order to measure 
their phase-lags. About half of the sample is composed of sources for 
which phase-lags were determined 
in the 1980's, 
the ones for which both works are in agreement serving as benchmark objects 
while for the objects with clear discrepancy the aim being re-determination 
of their phase-lag. The rest of the sample consists of objects with no 
recorded phase-lag measurements. 
About half of the sources of the sample have been previously imaged but the 
interferometric observations were taken at a random time and/or with poor 
sensitivity. 
We are currently in a process of imaging all the sources in the sample 
with either eMERLIN or JVLA around the OH maxima of each source, as 
predicted from the NRT light curves, in order to improve the angular diameter 
determination by detecting the faint interpeak signal and better constrain the 
shell (a)symmetries. \\

Past reports of the method and status of the project were presented in 
\cite[Engels \etal\ (2012, 2015)]{Engels15a} while a detailed description of 
the applicability of the method in measuring distances for objects beyond the 
solar vicinity 
can be found in \cite[Etoka \etal\ (2014)]{Etoka_etal14}. 
In the next section an update and discussion based on the results obtained 
so far is presented.

\section{Discussion}\label{sec: Discussion}

\begin{table}[ht]
\begin{minipage}{0.33\linewidth}
  \caption{Summary of the periods, phase-lags, linear \& angular 
           diameters and distances inferred for the entire sample}
  \begin{center}
  {\scriptsize
    \begin{tabular}{|llrr|}
\hline\noalign{\smallskip}
                &              & min.      & max.        \\
\noalign{\smallskip}\hline\noalign{\smallskip}
P$^{a}$          & [yrs]:       & 1.16      & 6.05        \\ 
$\tau_0$$^{a}$   & [days]:      & $<10$     & 110         \\
2\,R$_{OH}$$^{a}$ & [$10^3$ AU]: & $<1.7$    & 19          \\
\noalign{\smallskip}\hline
$\phi$     & ["]:         & 0.8$^{b}$  &  8.0$^{b}$  \\
D          & [kpc]:       & 0.5$^{b}$  & 10.6$^{b}$  \\
\noalign{\smallskip}\hline
    \end{tabular} \\
  }
  \end{center}
\label{Table: Overall sample summary}
\end{minipage}\hfill
\begin{minipage}{0.65\linewidth}
 \caption{Results for OH~83.4-0.9 \& OH~16.1-0.3}
  \begin{center}
   {\scriptsize
     \begin{tabular}{|lrrrrrrrrrr|}
\hline\noalign{\smallskip}
  Object    &   &  P     &  & $\tau_0$ & & 2\,R$_{OH}$  &  &  $\phi$    &  &  D         \\
            &   & [yrs]  &  & [days]   & & [$10^3$ AU] &  &  ["]       &  & [kpc]      \\
\noalign{\smallskip}\hline\noalign{\smallskip}
OH 83.4-0.9 &  & 4.11    &  & 30       & & 5.2         &  & $\sim$1.8  &  & $\sim$3.0   \\ 
OH 16.1-0.3 &  & 6.03    &  & 110      & & 19.0        &  & $\sim$3.5  &  & $\sim$5.5   \\
\noalign{\smallskip}\hline
     \end{tabular}
   }
  \end{center}
 \vspace*{+0.2cm}
{\small
\noindent
{\bf a:} All the periods, phase lags and corresponding linear diameters are 
inferred from our NRT  monitoring. The status of which can be followed 
here: \texttt{http://www.hs.uni-hamburg.de/nrt-monitoring} \\
\noindent
{\bf b:} from the literature 
}
\label{Table: OH83 and OH16}
\end{minipage}
\end{table}

Table~\ref{Table: Overall sample summary} presents the summary of the results 
obtained so far. The first half of the table gives the range of periods, 
phase-lags and linear diameters inferred from the NRT monitoring from all 
the sources of the sample. The second half of the table gives the range of 
angular diameters and inferred phase-lag distances.
The phase-lags measured account for linear shell diameter of $\sim$1700 to 
$\sim$19000~AU. Generally, the diameter is larger for longer-period objects. 
Distances ranges from 0.5 to 10.6~kpc. But, it has to be noted, that at 
the time of writing, although 70\% of the sources of the sample have a 
measured angular diameters, 60\% are from the literature including these 
2 extremas. \\

\begin{figure}[ht]
\begin{center}
 \includegraphics[width=13.5cm]{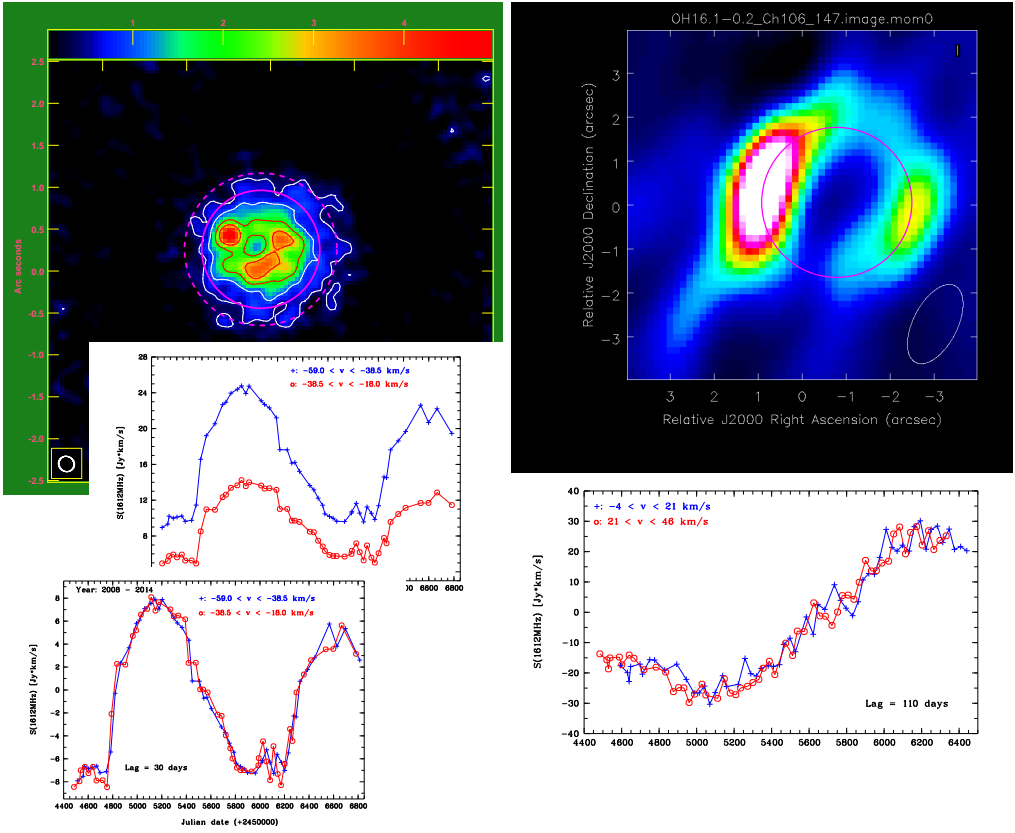} 
\caption{{\em Left main-panel:} eMERLIN map of the integrated emission 
over the inner part of the spectrum of OH~83.4-0.9 covering the velocity 
range ∼[$-55$;$-23$]~km~s$^{-1}$. The magenta full and dotted circles delineate 
the outer part of the shell.  {\em main-panel insert:} spectrum. 
{\em middle:} ``raw'' blue-peak and red-peak light curves of OH 83.4-0.9 
obtained from the NRT monitoring over a period of 7 years. {\em bottom:} 
scaled and shifted light curves for phase-lag determination. 
{\em Right main-panel:} JVLA map of the integrated emission over the red 
inner part of the spectrum of OH 16.1−0.3 covering the velocity range 
[+27; +37]~km~s$^{-1}$. 
The magenta circle presents the best fit of the projected diameter of the shell.
{\em main-panel insert:} Spectrum. 
{\em bottom:} scaled and shifted light curves for phase-lag determination. 
}
\label{fig: Concrete examples}
\end{center}
\end{figure}

Figure~\ref{fig: Concrete examples} presents the NRT monitoring and 
interferometric mapping for OH~83.4-0.9 and OH~16.1-0.3, two objects of the 
sample for which there was no previous imaging. We mapped both objects around 
their OH maximum with eMERLIN and the JVLA respectively, which allowed us to 
retrieve a substantial amount of the faint interpeak emission. For both 
objects, the channel maps obtained are in agreement with the shells being 
spherically-thin in uniform radial expansion. As an illustration of the 
phase-lag determination method explained in Section~\ref{sec: Method}, the 
left middle- and bottom-panels show the ``raw'' blue-peak and red-peak light 
curves of OH~83.4-0.9 and the scaled and shifted light curves leading to the 
phase-lag measurement.
The period, phase lags and corresponding linear \& angular diameters measured 
and subsequent inferred distances for these 2 objects are summarized 
in Table~\ref{Table: OH83 and OH16}. 
For these 2 objects, we estimated the uncertainty of the linear diameter to be 
within 10\%, while that of the angular diameter to be within 15\%, leading to a 
distance determination uncertainty of less than 20\%.
On the other hand, the distance determinations given in 
Table~\ref{Table: Overall sample summary} are still questionable as strongly 
dependent on the degree of exploration for the shell extent determination, 
i.e., faint tangential emission, which not only allow to better constrain the 
actual extent of the shell, but also its actual geometry, as in particular, 
a strong deviation from a spherically thin shell in uniform radial expansion 
can lead to distance uncertainty greater than 20\% (Etoka \& Diamond, 2010).

\section{Closing Notes}
The main contribution to the early phase-lag inconsistencies could be due to: 
incomplete coverage of lightcurves; inhomogeneous sampling; use of analytical 
functions to fit the lightcurves.
Phase-lag distances can be determined with an uncertainty of less 
than $20$\%, provided that a good constraint on both the linear and 
angular diameter determinations can be achieved. The main factors for 
doing so are the following: 
\begin{itemize}
 \item the shape of the light curves must be well defined. 
       This can be obtained with high cadence monitoring observations 
       (i.e., typically with ~0.03 P); 
 \item the light curves cover more than one period;
 \item the faint tangential emission tracing the actual full extent 
       of the shell can be imaged via high-sensitivity interferometric 
       observations better retrieved around the maximum of the OH cycle;
 \item significant shell asymmetries can be excluded or modelled.  
\end{itemize}

\end{document}